\documentclass[aps,prl,twocolumn,noshowpacs,superscriptaddress,groupedaddress, floatfix]{revtex4}  
\usepackage{graphicx}  
\usepackage{dcolumn}   
\usepackage{bm}        
\usepackage{amssymb}   
\usepackage{amsmath}
\usepackage[dvipsnames]{xcolor} 
\usepackage[colorlinks=true, pdfstartview=FitV, linkcolor=blue, citecolor=blue, urlcolor=blue]{hyperref} 
\usepackage{epstopdf}
\usepackage[free-standing-units=true, per-mode=symbol-or-fraction, range-phrase={-}, separate-uncertainty=true]{siunitx} 
\DeclareSIUnit{\sqrthz}{\ensuremath{\sqrt{\text{\hertz}}}} 
\DeclareMathOperator{\sign}{sign}
\usepackage{algorithm}
\usepackage{algpseudocode}

\newcommand{\seeAppendix}[2]{Appendix~\hyperref[#2]{#1}}
\newcommand{\circa}{ca.\@}

\newcommand{\affilIFP}{Laboratory for Solid State Physics, ETH Zurich, CH-8093 Zurich, Switzerland.}
\newcommand{\affilJEOL}{JEOL RESONANCE Inc., Musashino, Akishima, Tokyo 196-8558, Japan.}
\newcommand{\affilBristol}{School of Cellular and Molecular Medicine, Faculty of Life Sciences, Biomedical Sciences Building, University of Bristol, Bristol BS8 1TD, UK.}
\newcommand{\affilUZ}{Institute of Molecular Life Sciences, University of Zurich, Winterthurerstrasse 190, CH-8057, Zurich, Switzerland.}

\begin{document}

\preprint{}

\title{Force-detected Magnetic Resonance Imaging of Influenza Viruses in the Overcoupled Sensor Regime}

\author{Marc-Dominik Krass}\affiliation{\affilIFP}
\author{Nils Prumbaum}\affiliation{\affilIFP}
\author{Raphael Pachlatko}\affiliation{\affilIFP}
\author{Urs Grob}\affiliation{\affilIFP}
\author{Hiroki Takahashi}\altaffiliation[Current affiliation: ]{\affilJEOL}\affiliation{\affilIFP}
\author{Yohei Yamauchi}\affiliation{\affilBristol}\affiliation{\affilUZ}
\author{Christian L. Degen}\affiliation{\affilIFP}
\author{Alexander Eichler}\affiliation{\affilIFP}

\date{\today}


\begin{abstract}
Long and thin scanning force cantilevers are sensitive to small forces, but also vulnerable to detrimental non-contact interactions. Here we present an experiment with a cantilever whose spring constant and static deflection are dominated by the interaction between the tip and the surface, a regime that we refer to as ``overcoupled''. The interactions are an obstacle for ultrasensitive measurements like nanoscale magnetic resonance imaging (nanoMRI). We discuss several strategies to overcome the challenges presented by the overcoupling, and demonstrate proton nanoMRI measurements of individual influenza virus particles.
\end{abstract}


\maketitle



\paragraph{Introduction} - Scanning force microscopy (SFM) is a wide and mature field that finds applications in physics, biology, chemistry, as well as the materials and surface sciences~\cite{Binnig_1986, jalili_2004review, kazakova_2019frontiers, Giessibl_2003, krieg_2019atomic}. State-of-the-art SFM can attain atomically resolved surface scans or follow nanoscale processes with video rates, making these instruments invaluable for both academic and industrial research. An open target of the community is to reach the zeptonewton force range, which is expected to enable the detection of individual nuclear spins. This is a crucial precondition for magnetic resonance force microscopy~\cite{sidles_1991noninductive, rugar_2004single, Degen_2009} with near-atomic spatial resolution, with exciting prospects in structural biology and for the characterization of quantum devices.

In order to realize zeptonewton force detection, the internal damping and stiffness of nanomechanical sensors must be reduced as much as possible. Many groups are striving to optimize different devices as force sensors, including cantilevers~\cite{mamin_2001sub, tao_2014single, heritier_nanoladder_2018}, doubly-clamped beams~\cite{Verbridge_2008, Anetsberger_2010, Ghadimi_2018, Beccari_2021, beccari_2021hierarchical, gisler2021soft}, membranes~\cite{Reinhardt_2016, Tsaturyan_2017, Reetz_2019, Halg_2021, seis_2021ground}, nanowires~\cite{Nichol_2012, rossi_2017vectorial, deLepinay_2017universal, sahafi_2019ultralow}, graphene sheets~\cite{weber2016force}, carbon nanotubes~\cite{moser_2013ultrasensitive, deBonis_2018ultrasensitive}, and levitated particles~\cite{hempston_2017force, Hebestreit_2018}. The main challenge, in most cases, is to preserve the excellent internal characteristics of a force sensor in close proximity to a sample. When the distance between a sensor and the sample is sufficiently reduced, non-contact forces set in, providing an external source for damping (dissipative forces) and stiffness (spring forces). For sensors with extremely low intrinsic spring or damping forces, the non-contact forces can dominate the frequency and linewidth of a sensor. In this case, we speak of an ``overcoupled'' resonator, in analogy to the terminology used for superconducting circuits and optical cavities~\cite{Circuit_quantum_electrodynamics,aspelmeyer_cavity_2014}.

Dissipative non-contact forces have been the subject of intense research efforts for more than two decades~\cite{stipe_noncontact_2001, volokitin_noncontact_2003, zurita-sanchez_friction_2004, kuehn_dielectric_2006, volokitin_near-field_2007, yazdanian_dielectric_2008, kisiel_suppression_2011, she_noncontact_2012, den_haan_spin-mediated_2015, de_voogd_dissipation_2017, Heritier_2021}. By reducing the quality factor of a resonator, dissipation generally impedes a sensor's ability to detect small forces. Conservative non-contact forces, by contrast, act in phase with the oscillation and do not produce damping. These forces are responsible for changes of the sensor frequency close to a sample surface, which is the basis of a detection method used in high-resolution scanning force experiments~\cite{albrecht_1991frequency,giessibl_1995}. In the overcoupled regime, the energy potential created by conservative non-contact forces can overwhelm the intrinsic spring force of the resonator. As a consequence, the resonator experiences large static deflections, frequency shifts that exceed its natural resonance frequency, and even instabilities at certain positions. 
While such effects are mostly observed at tip-surface distances up to a few nanometers in most SFM setups~\cite{uchihashi_1997}, they can already become a serious obstacle at larger distances for resonators with very low spring constants. Conservative non-contact forces therefore deserve to be considered for the next generations of ultrasensitive probes.

In this paper, we experimentally investigate the impact of conservative non-contact forces in the overcoupled regime. Our measurement setup is a Magnetic Resonance Force Microscope (MRFM)~\cite{grob_magnetic_2019}, which we use to detect ensembles of nuclear spins and to perform three-dimensional nanoscale magnetic resonance imaging. Because of the small spring constant of the cantilever, the apparatus is highly sensitive to the influence of static forces. We demonstrate how conservative non-contact forces impact our measurements and how their detrimental effect can be partially mitigated with various strategies. As a result, we are able to reconstruct nanoMRI images of individual influenza viruses from our data. We further show that our cantilever sensor can be used as a static probe for surface forces with a sensitivity of $\SI{30}{\femto\newton}$.

\begin{figure}[htb]
\includegraphics[width=\columnwidth]{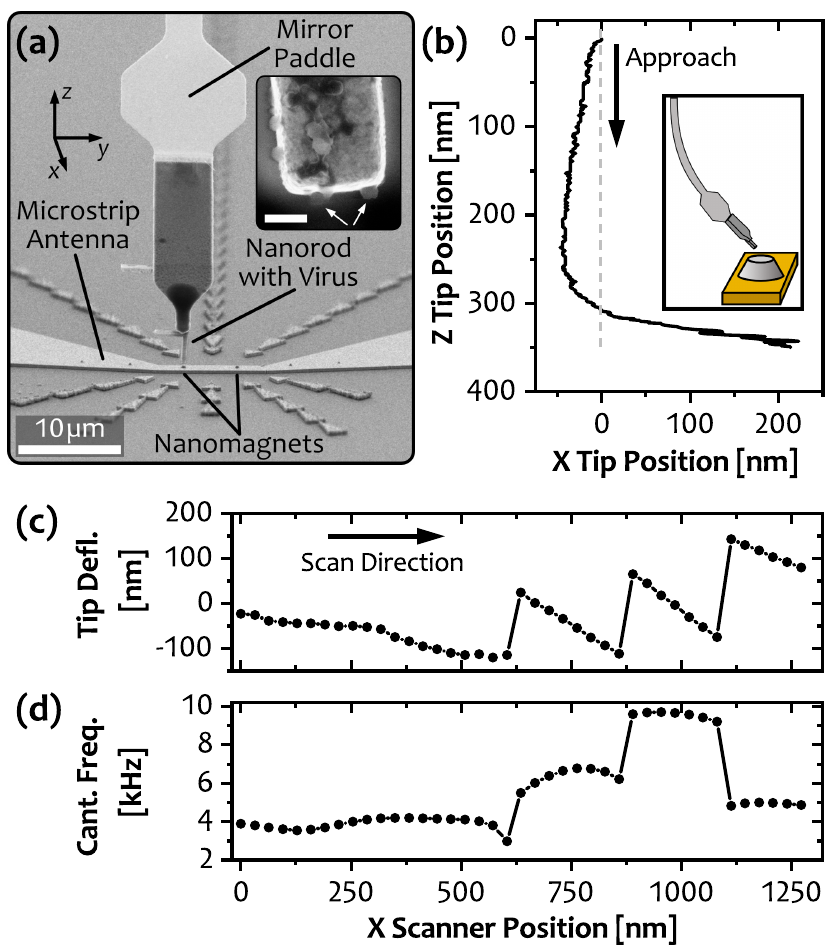}
\caption{Static cantilever deflection. \textbf{(a)}~Illustration of the spatial arrangement of the cantilever apex positioned $\SIrange{10}{100}{\nano\metre}$ above a nanomagnet that is fabricated on top of a microstrip antenna. The image is composed of two individual SEM micrographs taken of the cantilever and the antenna chip. Inset: image of a nanorod apex (a different device was used in this study). Individual virus particles are marked by the arrows. The scale bar has a length of $\SI{200}{\nano\metre}$.  \textbf{(b)}~Cantilever tip position during an approach curve next to the nanomagnet. Inset: schematics of the bent cantilever (not to scale). We define the cantilever deflection as the difference between the tip position and the scanner position, which is set to $a_\mathrm{cant} = \SI{0}{\nano\metre}$ at the starting point in this case. \textbf{(c)}~Cantilever deflection and \textbf{(d)}~cantilever frequency during an MRFM scan across the microstrip antenna at a cantilever-nanomagnet distance of $\SI{10}{\nano\metre}$. Points indicate the measurement positions. In this measurement, a virus particle was centered above the nanomagnet at $x = \SI{910}{\nano\metre}$ and interacting strongly with it between $x \approx \SI{650}{\nano\metre}$ and \SI{1165}{\nano\metre}.}
\label{fig:fig1}
\end{figure}


\section{Setup}
Our sensor is a silicon cantilever with dimensions $\SI{150}{\micro\meter} \times \SI{4}{\micro\meter} \times \SI{130}{\nano\meter}$, a bare resonance frequency of $f_0 = \SI{3500}{\hertz}$, a mass of $m = \SI{e-13}{\kilo\gram}$, and a quality factor of $Q = 25000$ in vacuum and at the base temperature $T = \SI{4.7}{\kelvin}$ of our helium bath cryostat. These numbers translate into an intrinsic spring constant of $k_{0} = \SI{50}{\micro\newton\per\metre}$ and a friction coefficient of $\gamma_{0} = \SI{e-13}{\kilo\gram\per\second}$ . The sensor displacement is detected via a laser interferometer with a power of $\SIrange{10}{100}{\nano\watt}$ directed onto a paddle near the cantilever tip. The interferometer has a displacement detection sensitivity of roughly $\SI{2e-12}{\metre\per\sqrthz}$ around the cantilever resonance frequency. Importantly, the interferometer can also be used to monitor slow deflections of the cantilever tip through the total reflected power~\cite{grob_magnetic_2019, krass22phd}. The detection uncertainty of this method is limited by temperature drift of the laser, which induces wavelength changes that produce the same signature as mechanical motion. By controlling the laser temperature actively with a precision of \circa{} $\SI{10}{\milli\kelvin}$, we achieve a root-mean-square uncertainty of about $\SI{0.2}{\nano\metre}$ in a $\mathrm{DC}-\SI{500}{\hertz}$ bandwidth.

To attach the influenza virus particles to the cantilever, we use a two-step approach. First, a silicon chip containing cleanroom-fabricated nanorods with a bottom cross section of \circa{} $\SI{500}{\nano\metre} \times \SI{500}{\nano\metre}$ is dipped into a low-concentration virus suspension. The concentration is adjusted to result in a small number of well-separated virus particles on the bottom surface of the nanorod. After a chemical fixation step, a single nanorod is glued manually to the end of a cantilever. This process conserves the three-dimensional structure of the virus and does not degrade the mechanical properties of the cantilever~\cite{overweg2015probing}.

Figure~\ref{fig:fig1}(a) illustrates the spatial arrangement of the cantilever in an MRFM experiment. The cantilever is mounted in a pendulum geometry to avoid vertical snap-into-contact. For nanoMRI measurements, the cantilever apex is positioned $\SIrange{10}{150}{\nano\metre}$ above the top surface of a nanomagnet, which is a truncated cone fabricated on top of a microstrip antenna~\cite{poggio2007nuclear}. The nanomagnet is made of FeCo, has a top/bottom diameter and a height of $d_\mathrm{top} = \SI{300}{\nano\metre}$, $d_\mathrm{bottom} = \SI{500}{\nano\metre}$, and $h_\mathrm{mag} = \SI{200}{\nano\metre}$, respectively, and an estimated saturation magnetization of $\mu_{0}M_\mathrm{sat} = \SI{1.88}{\tesla}$. This nanomagnet provides the magnetic field gradient $G$ that generates the measured spin-mechanics coupling~\cite{grob_magnetic_2019}.

As the cantilever approaches the nanomagnet, it experiences strong conservative non-contact forces that bend its tip away from the nominal position, cf.\@ Fig.~\ref{fig:fig1}(b). This deflection is caused by inhomogeneous lateral tip-surface forces. In Fig.~\ref{fig:fig1}(c) and (d), we show discontinuities of the cantilever's position and frequency due to these forces during a sweep along the $x$ direction (at constant height $z$). At each discontinuity, the cantilever position changes abruptly by more than \SI{100}{\nano\meter} and the frequency $f_\mathrm{cant}$ jumps by several $\si{\kilo\hertz}$. The maximum frequency of \circa{} $\SI{10}{\kilo\hertz}$ at $x_\mathrm{scan} = \SI{1000}{\nano\metre}$ is almost three times larger than $f_0$. Similar patterns are typically observed with lateral force microscopy in the regime of atomic stick-slip motion, however on a much smaller length scale (on the order of $\SI{100}{\pico\metre}$ \cite{jansen_2010} rather than $\SI{100}{\nano\metre}$). Dissipative non-contact forces, which we also observe, are shown in the \seeAppendix{A}{APP:forces}.


\begin{figure}[htb]
\includegraphics[width=\columnwidth]{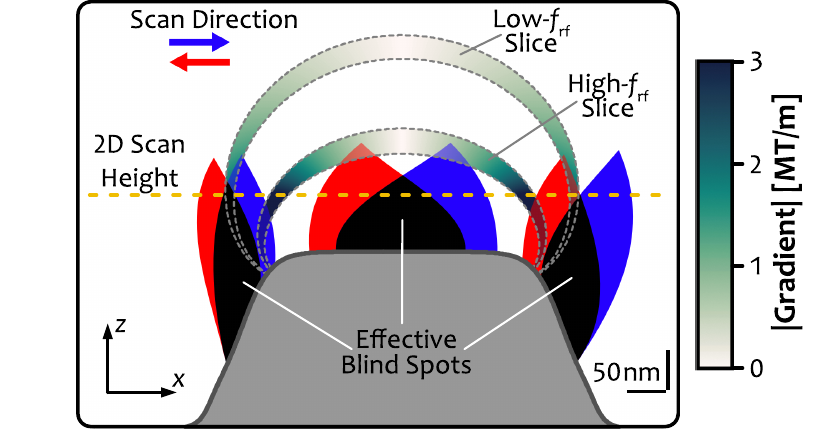}
\caption{Schematic illustration of the nanomagnet, two resonant slices, and the position of blind spots for positive (blue) and negative (red) scan directions in $x$. The color coding inside the resonance slices indicates the absolute value of the magnetic field gradient $G$ generated by the nanomagnet (light gray), see \seeAppendix{C}{APP:magnetic_simulations} for details. The dashed yellow line indicates the height of a typical $x-y$ scan with various blind spots. The tip-surface interactions decrease with increasing separation $\Delta z$ and the blind spots vanish for $\Delta z \gtrsim \SI{100}{\nano\metre}$.
}
\label{fig:fig3}
\end{figure}

\section{MRFM Case Study}
In MRFM, spins are coupled to the cantilever via a nanomagnetic field gradient~\cite{sidles_1991noninductive}. To create a force that acts resonantly on the cantilever, the spins are periodically inverted via radio-frequency pulses~\cite{Degen_2009,grob_magnetic_2019}. Spatial resolution is achieved by the condition that the spin's Larmor frequency must match the pulse carrier frequency $f_\mathrm{rf}$ in order to allow inversions. This condition defines ``resonant slices'' in space where the inversion protocol is effective. The width of a slice is controlled by the frequency modulation depth $\Delta f_\mathrm{rf}$ of the pulse, which extends the resonance condition to a range of Larmor frequencies $f_\mathrm{rf} \pm \Delta f_\mathrm{rf}$. Two such slices are illustrated in Fig.~\ref{fig:fig3}, corresponding to different carrier frequencies $f_\mathrm{rf}$. Nuclear spins at different positions inside the sample are brought into the slice for different scanner positions, and the force measured at each position is used as a measure to count spins. Since the acquired signal is a convolution of the spin density in the sample with the slice geometry, a reconstruction step is necessary to obtain the true sample structure~\cite{Degen_2009}.

\begin{figure*}[htb]
\includegraphics[width=\textwidth]{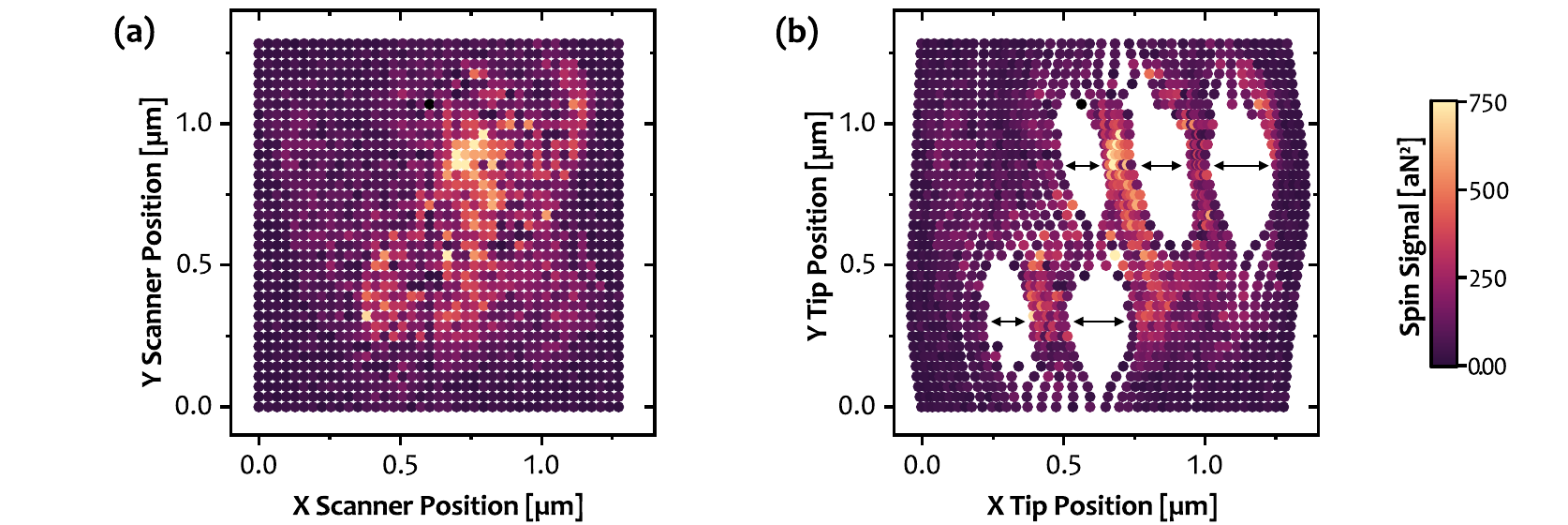}
\caption{MRFM measurements in the overcoupled sensor regime.
\textbf{(a)}~A two-dimensional MRFM scan recorded as a function of $x$ and $y$ position of the scanning stage on a regular grid with a step size of ca.\@ $\SI{30}{\nano\metre}$. We used $f_\mathrm{rf} = \SI{255.5}{\mega\hertz}$ and $\Delta f_\mathrm{rf} = \SI{800}{\kilo\hertz}$ at a cantilever-nanomagnet separation of $\SI{30}{\nano\metre}$ to detect \textsuperscript{1}H spins (see main text for explanation). The position of the nanomagnet was approximately in the center of the field of view. \textbf{(b)}~Same data as in panel~(a), but plotted as a function of the actual cantilever tip position. Arrows indicate the opening of the blind spots. To obtain the tip position, the scanner position is corrected by the static cantilever deflection. Note that the $y$ position of the tip is not corrected, as the cantilever is much stiffer in this direction and static deflections are assumed to be negligible.}
\label{fig:fig2}
\end{figure*}

To visualize the impact of conservative non-contact forces on our MRFM measurements, we show an example of a two-dimensional MRFM scan at constant $z$ position in Fig.~\ref{fig:fig2}(a). Each data point corresponds to a force measurement at a certain scanner position. Bright (dark) data points indicate that a large (small) number of hydrogen atoms is located inside the resonant slice and contributes to the spin signal. This image, however, does not reflect the true spatial distribution of the signal: in Fig.~\ref{fig:fig2}(b), we show the same data after correcting for the static tip deflection along $x$ measured with the interferometer. We observe that the cantilever is pushed away from certain regions by conservative non-contact forces, creating ``blind spots'' in real space where no signal can be measured. The spots appear around the positions where we expect maximum signal, depriving us of crucial information.

Figure~\ref{fig:fig2} reveals a fundamental problem arising with ultrasensitive, overcoupled sensors: to achieve the desired sensitivity to small forces, our cantilever is designed to have a very low spring constant. This, however, makes it also sensitive to arbitrary surface forces that create instabilities and obstruct the collection of data in certain regions of space. We can at present not identify the microscopic origin of these forces with certainty, but because of their long-range characteristics, it is likely that they are related to trapped charges at the interface between the nanomagnet and the microstrip antenna~\cite{Heritier_2021}.
In the following, we discuss three strategies to overcome the problems arising from the blind spots, and to perform nanoscale MRI even in the presence of strong non-contact forces. The strategies are illustrated in Fig.~\ref{fig:fig3}. Example plots showcasing the results for our instrument are collected in the \seeAppendix{B}{APP:imaging_strategies}.

\paragraph{Strategy~(i): increasing the tip-surface distance -} non-contact forces usually drop off quickly as a function of the distance between tip and surface. In our case, increasing the height over the nanomagnet surface from \SI{30}{\nano\meter} to \SI{100}{\nano\meter} leads to significantly smaller blind spots. At a separation of \SI{150}{\nano\meter}, the influence of conservative non-contact interactions is negligible. In Fig.~\ref{fig:fig3}, this effect is schematically shown as a gradual closing of the blue (red) petal-shaped blind-spot regions when scanning in positive (negative) $x$ direction. In many experiments, an increased separation will also somewhat reduce the ability of the sensor to resolve small forces of interest, e.g.\@ small nuclear spin forces.

\begin{figure*}[htb]
\includegraphics[width=\textwidth]{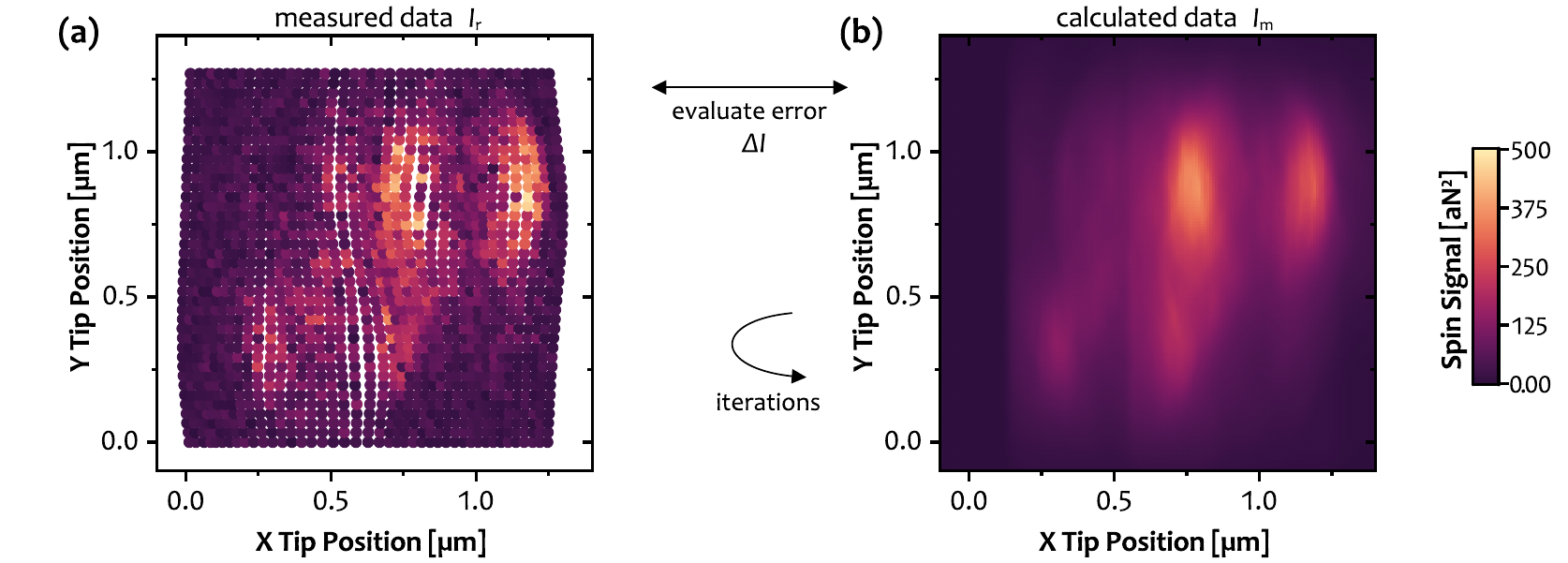}
\caption{Iterative reconstruction process. \textbf{(a)}~Measured MRFM data
$I_\mathrm{m}$ recorded at a cantilever-nanomagnet distance of 
$\Delta z = \SI{110}{\nano\metre}$. For the reconstruction, the cantilever
tip positions are coerced to nearest-neighbor positions on a 
$10\times 10\times 10$\,nm grid. \textbf{(b)}~Example of the calculated MRFM signal $I_\mathrm{r}$ at $\Delta z = \SI{110}{\nano\metre}$ after a successful
reconstruction with three $x-y$ planes ($\SI{110}{\nano\metre}$, $\SI{130}{\nano\metre}$, $\SI{150}{\nano\metre}$). During each iteration, the reconstruction algorithm
reduces the error $\Delta I = I_\mathrm{m}-I_\mathrm{r}$ between both 
data sets until it drops below a threshold.
}
\label{fig:fig4}
\end{figure*}

\paragraph{Strategy~(ii): combining different scan directions -} in general, the blind spots for scans in positive and negative scan directions do not overlap fully, cf.\@ blue and red petals in Fig.~\ref{fig:fig3}. This hysteresis can be exploited by scanning in both directions and combining the results in one data set. As a result of this strategy, only the overlap regions shown in black remain inaccessible.

\begin{figure}[htb]
\includegraphics[width=\columnwidth]{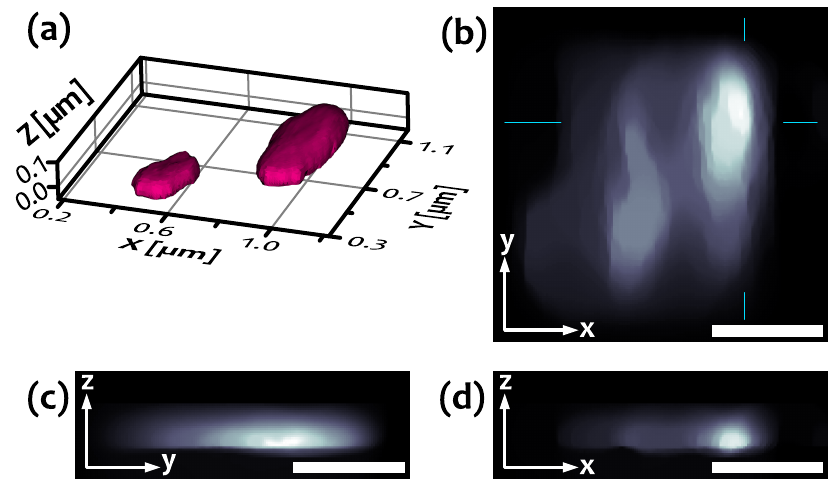}
\caption{Reconstructed nanoMRI image $O_\mathrm{r}$ of influenza viruses. \textbf{(a)}~Isosurface image of the three-dimensional hydrogen nuclear spin density calculated from the MRFM data. The image shows two regions of high spin density whose sizes agree with an interpretation of a single virus (left) and an aggregation of 2-3 viruses (right). The three monochromatic images show hydrogen nuclear spin density in planes at \textbf{(b)}~$z = \SI{80}{\nano\meter}$, \textbf{(c)}~$x = \SI{1}{\micro\meter}$, and \textbf{(d)}~$y = \SI{0.9}{\micro\meter}$. The blue lines in panel~(b) indicate the positions of the planes shown in panel~(c)/(d). All scale bars are $\SI{500}{\nano\meter}$ long. Bright (dark) color scale corresponds to high (low) measured spin density.}
\label{fig:fig5}
\end{figure}

\paragraph{Strategy~(iii): changing the shape of the PSF -} in MRFM, the fact that the spin signal is not collected from a single point but from an extended point-spread function (PSF) opens a third possibility to avoid the blind spots. On the one hand, any spin inside the sample can be detected at various tip positions that place the spin in different parts of an active resonant slice. On the other hand, different resonant slices can be activated by changing the carrier frequency. For the example shown in Fig.~\ref{fig:fig3}, while the regions of strong signal, i.e.\@ large gradient $G$ (dark green), are inaccessible with the low-$f_\mathrm{rf}$ slice because of the blind spots, one half of the high-$f_\mathrm{rf}$ slice is clear of blind spots for either scan direction.
This strategy, or variants thereof with several slices used simultaneously~\cite{moores_2015accelerated}, only applies to scanning probe microscopy methods whose PSFs are not concentrated at a single point.



\section{Application and Result}
In our MRFM experiment, we selected strategy~(i) to obtain an optimal compromise between high magnetic field gradient and small blind spots. Strategies (ii) and (iii) are potentially more powerful, but the reconstruction algorithm we currently use cannot efficiently handle the large differences in the point-to-point separations. Further, we chose the $y$-axis as fast scanning direction in order to reduce the number of discontinuities in one line and the mechanical hysteresis, cf.\@ Fig.~\ref{fig:fig1}(c) and (d) and Fig.~\ref{fig:figS2}, respectively.

The blind spots decrease significantly at distances greater than $\SI{100}{\nano\metre}$ above the magnet surface, leading to a maximum static deflection of less than $d_\mathrm{max} = \SI{70}{\nano\metre}$ and a mean deflection of $d_\mathrm{mean} = \SI{30+-16}{\nano\metre}$, compared to $d_\mathrm{max, 10nm} = \SI{170}{\nano\metre}$ and $d_\mathrm{mean, 10nm} = \SI{61+-39}{\nano\metre}$ at a distance of $\SI{10}{\nano\metre}$. In Fig.~\ref{fig:fig4}, we show one out of three measured $x-y$ planes (at distances $\SI{110}{\nano\metre}$ to $\SI{150}{\nano\metre}$) that contained only small blind spots and thus were used for the reconstruction. 
Notice that the signal is not dramatically lower than for the data in Fig.~\ref{fig:fig2} because the magnetic gradient falls off roughly on the length scale of the magnet itself, which is about \SI{250}{\nano\meter}.

The measured image $I_{\mathrm{m}}$ does not directly correspond to the spin density $O$ of the attached sample. The measured signal at one measurement position is given by the convolution of $O$ with the PSF, which in turn is controlled by the nanomagnet shape and the pulse parameters. In our image-formation model, we write
\begin{equation}
     I_{\mathrm{m}} = \mathit{PSF}*O\,,
\end{equation}
where $I_{\mathrm{m}}$ is the measured image. To recover the object structure from the measured data, we use a regularized least-squares optimization. The approach we chose, along with a possible algorithm, is described in the \seeAppendix{D}{APP:ADMM}.

Figure~\ref{fig:fig5} shows the reconstructed nanoMRI image obtained from our experiment. In spite of the challenges presented by the strong non-contact interaction, the reconstructed image clearly shows two regions of large nuclear spin signal. The small and large blobs possibly correspond to a single virus and the aggregation of 2-3 viruses, respectively. The nanoMRI image agrees qualitatively with a topography scan performed in our setup, cf.\@ \seeAppendix{E}{APP:topography}. We conclude that our strategy is successful and that the influence of conservative non-contact forces can be mitigated to some degree. Nevertheless, the spatial resolution of the image in Fig.~\ref{fig:fig5} is affected by the fact that we were not able to include the blind spot-corrupted data from the highest magnetic field gradient regions into the reconstruction.
 
While static deflection presented a problem for our MRFM experiments, it can potentially be a useful probe in some surface studies. With a root-mean-square position uncertainty of \SI{0.2}{\nano\meter} and a spring constant of \SI{50}{\micro\newton\per\metre}, our cantilever is sensitive to static forces down to \SI{30}{\femto\newton}, which is close to the performance of other force sensors read out at much higher laser powers \cite{Rugar_1989, smith_2009}.

\section{Outlook}
Conservative non-contact forces will become increasingly important in future generations of nanomechanical scanning probes with optimized sensitivity. 
Operating such devices in the overcoupled regime changes the resonance frequency, alters the mode structure, and, specifically for MRFM, can dramatically reduce the inversion pulse efficiency. In combination, these effects can significantly reduce the resonator's force sensitivity.
It is therefore important to be aware of this potential issue, monitoring the relevant signatures (like large frequency shifts), and implementing counter-strategies when necessary. Our work outlines the basic problem and demonstrates how it can be handled even in a very complex type of experiment. 

\section*{Acknowledgments}

This work was supported by Swiss National Science Foundation (SNFS) through the National Center of Competence in Research in Quantum Science and Technology (NCCR QSIT) and the Sinergia grant (CRSII5\_177198/1), and the European Research Council through the ERC Starting Grant ``NANOMRI'' (Grant agreement 309301). Sample fabrication was carried out in the FIRST cleanroom at ETH Zurich and the BRNC cleanroom facility at IBM Ruschlikon. We thank Jan Rhensius for help with the sample attachment process.

\appendix
\section{Appendix A: Map of conservative and dissipative interactions} \label{APP:forces}

In the main text, we focus on the effect of conservative non-contact forces, i.e., a spring force generated by the proximity of the tip to the surface. However, we also observe dissipative non-contact forces which increase the effective damping constant of the cantilever. We extract the effective damping constant directly from the autocorrelation constant of time traces used for the spin detection. In this way, we gain access to the damping information without additional measurements.

\begin{figure}[htb]
\includegraphics[width=\columnwidth]{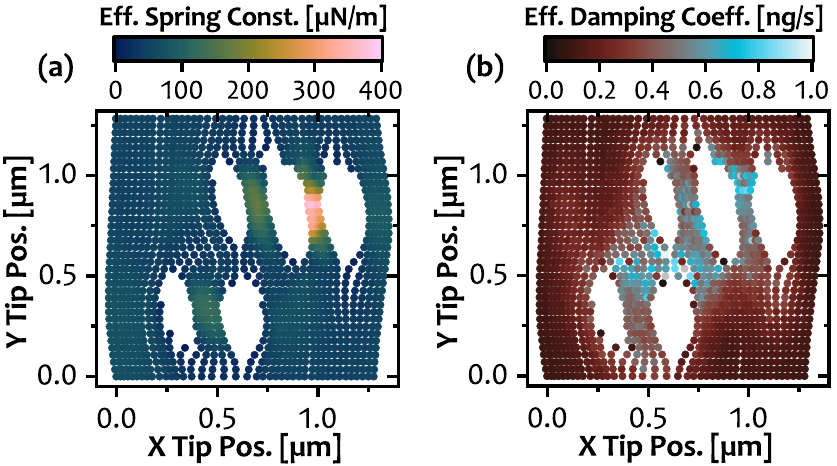}
\caption{Dependency of \textbf{(a)}~effective spring constant and \textbf{(b)}~effective damping coefficient on $x$ and $y$ position at $\Delta z = \SI{30}{\nano\meter}$. The data was obtained simultaneously with the MRFM signal shown in Fig.~\ref{fig:fig2}.}
\label{fig:figS1}
\end{figure}

The result for an example scan is shown in Fig.~\ref{fig:figS1}. We note that the numbers and the spatial distribution recovered here are different from those seen with a nanoladder cantilever over flat surfaces~\cite{Heritier_2021}. Here, the dissipative and conservative forces over the magnet vary over ranges of several \SI{100}{\nano\meter} and they have different spatial signatures. We also note that the non-contact damping coefficient is orders of magnitudes larger than in our previous experiments~\cite{Heritier_2021}, which is a result of the much larger apex of the cantilever used in this study.

\section{Appendix B: Imaging strategies} \label{APP:imaging_strategies}

\begin{figure}[h!]
\includegraphics[width=\columnwidth]{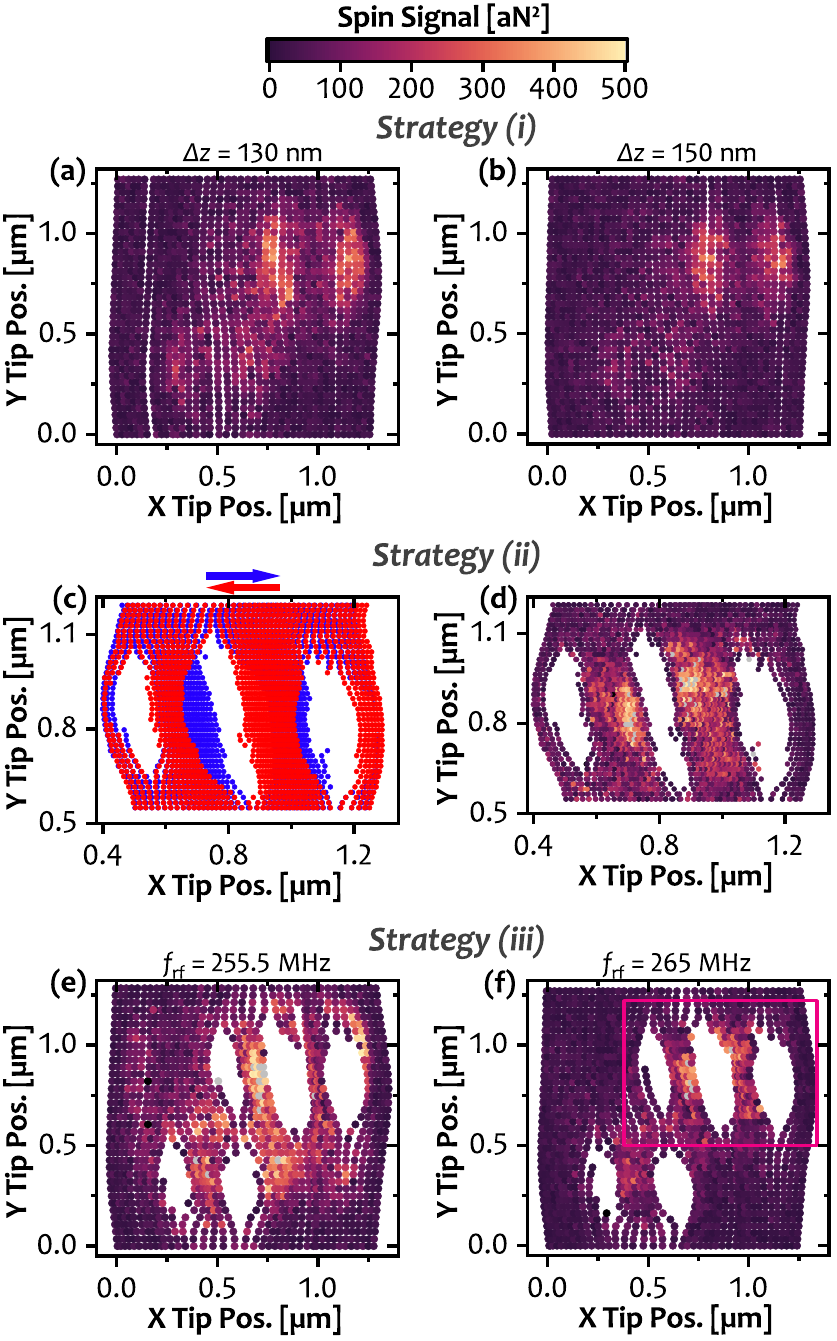}
\caption{\textbf{(a)}~and \textbf{(b)}: raw data of the two additional scans used for the reconstruction shown in Fig.~\ref{fig:fig5}. \textbf{(c)}~Demonstration of the cantilever tip position hysteresis between scans in opposite directions, shown in blue and red. This scan was performed at $\Delta z = \SI{40}{\nano\metre}$ on a smaller scan range than in panels (a) and (b). \textbf{(d)}~Spin signal image combined out of the two scans performed in opposite directions in (c). \textbf{(e)}~and \textbf{(f)}: scans performed at the same height $\Delta z =\SI{40}{\nano\metre} $ with different pulse central frequencies $f_\mathrm{rf}$, resulting in different information captured. Because of the smaller diameter of the resonant slice used for recording the data shown in panel~(f), the signal is more localized around the blind spots than in panel~(e), cf.\@ Fig.~\ref{fig:figS3} for a quantitative field simulation. The square frame in panel~(f) shows the measurement range of the zoom scan in panel~(d). The MRFM signal regions coincide without additional alignment, indicating the excellent mechanical stability of our setup.}
\label{fig:figS2}
\end{figure}

In Fig.~\ref{fig:figS2} we provide examples of all three strategies in our system. Strategy~(i) was selected for the reconstruction because it featured by far the smallest blind spots, which allowed for a significantly simpler reconstruction.

\section{Appendix C: Magnetic field simulation} \label{APP:magnetic_simulations}
\begin{figure}[htb]
\includegraphics[width=\columnwidth]{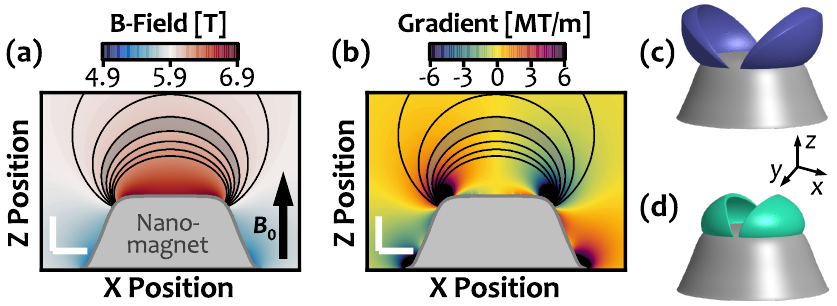}
\caption{\textbf{(a)}~Numerical simulation of the magnetic field $B_{0} + B_\mathrm{tip}(\vec{r})$ and \textbf{(b)}~the field gradient $G$ around the nanomagnet. The $B_{0}$-field of $\SI{5.88}{\tesla}$ points along the $z$-direction. The contour lines indicate regions of constant Larmor frequency (isochromats). The gray shaded area represents the shape of the resonant slice in which the spins are inverted by the inversion pulses. The scale bars have a length of $\SI{100}{\nano\metre}$. \textbf{(c)}~and \textbf{(d)}: Illustration of the shape of resonant slices in three dimensions. The colored contours indicate where the product between the magnetic field gradient and the PSF drops below a threshold value of \SI{0.5}{\mega\tesla\per\meter}. Simulations were performed for the measured magnet shape, an external field of $B_0 = \SI{5.88}{\tesla}$, a saturation magnetization of \SI{1.88}{\tesla} inside the magnet, and $f_\mathrm{rf} = \SI{255}{\mega\hertz}$ and $f_\mathrm{rf} = \SI{265}{\mega\hertz}$ in (c) and (d), respectively.}
\label{fig:figS3}
\end{figure}

In Fig.~\ref{fig:figS3} we show simulations of the nanomagnet properties and the resulting resonant slice geometries. The magnet's geometry was obtained from an AFM scan prior to installation into our MRFM apparatus. The calculations were performed by dividing the nanomagnet into thin disks with constant magnetization. The magnetic field of an individual disk can be found from an analytical expression~\cite{caciagli18}. Then, the sum of the magnetic fields generated by all disks is evaluated at each point in space~\cite{grob_magnetic_2019, krass22phd}.

\section{Appendix D: Algorithm to Solve Optimization Problem} \label{APP:ADMM}

The reconstruction problem expresses the task of finding the object $O_{\mathrm{r}}$ that is the optimal solution to
\begin{equation}\label{eq:1}
     \underset{O_{\mathrm{r}}}{\arg \min} \bigg( \frac{1}{2}\|I_{\mathrm{r}}-I_{\mathrm{m}}\|_{_2}^2 + \lambda \|O_{\mathrm{r}}\|_{_1} + \lambda_\mathrm{TV} \|D O_{\mathrm{r}}\|_{_2} \bigg)\ ,
\end{equation}
where $D$ is the isotropic total variation operator, $I_{\mathrm{m}}$ is the measured data, and $I_{\mathrm{r}}$ is the reconstructed image, which is defined as the three-dimensional convolution of the reconstructed object with the point-spread function.
The first term of the optimization problem in Eq.~\ref{eq:1} is minimized in order to reduce the quantitative difference between the measured data planes and the corresponding signal generated by the reconstruction model.
To further reduce the effect of measurement noise on the result, the two regularization terms are added. Selecting a large value for the weight $\lambda$ leads to sparser solutions and increasing $\lambda_\mathrm{TV}$ suppresses fast changes in the reconstructed object and thus reduces high-frequency noise. The chosen weights for the results in Fig.~\ref{fig:fig5} are $\lambda = 0.1$ and $\lambda_\mathrm{TV} = 7.5$. 

To solve the optimization problem in Eq.~\ref{eq:1}, it can be rewritten as
\begin{equation}
\begin{split}
    \min_{o_\mathrm{r},v,w} \bigg(\frac{1}{2} \|Po_\mathrm{r}-b_\mathrm{m}\|+\lambda\|v\|_{_1}+\lambda_\mathrm{TV}\|w\|_{_2} \bigg)\\
    \text{s.t.\quad} x -v = 0,\ \  Do_\mathrm{r}-w=0  \ .
    \end{split}
\end{equation}
This allows to use the framework developed by \textit{Gao et al.}~\cite{gao_admm} and the problem can be solved employing the alternating direction method of multipliers  (ADMM). In this work, the problem is solved using scaled dual variables, leading to the following update equations for the $(k+1)$-th iteration: 
\begin{align*} 
    o_{\mathrm{r},k+1} &= o_{\mathrm{r},k} - \alpha \Big[P^T \big(Po_{\mathrm{r},k}-b_\mathrm{m}\big)\\
    &\hphantom{{}=  o_{\mathrm{r},k} - \alpha \Big[  }+ \rho\big(o_{\mathrm{r},k}-v_{k}+\Tilde{v}_{k}\big)\\
    &\hphantom{{}=  o_{\mathrm{r},k}- \alpha \Big[  }+ \rho D^T\big(Do_{\mathrm{r},k}-w_k+\Tilde{w}_{k}\big)\Big]\\[0.5em]
    v_{k+1} &= \sign\big(o_{\mathrm{r},k+1}+\Tilde{v}_\mathrm{k}\big)\cdot\max\bigg(|o_{\mathrm{r},k+1}+\Tilde{v}_\mathrm{k}|-\frac{\lambda}{\rho},0\bigg)\\[0.5em]
    w_{k+1} &= \max\bigg(1-\frac{\lambda_\mathrm{TV}/\rho}{\|Do_{\mathrm{r},k+1}+\Tilde{w}_{k}\|_{_2}},0\bigg)\cdot\big(Do_{\mathrm{r},k+1}+\Tilde{w}_{k}\big)\\[0.5em]
    \Tilde{v}_{k+1} &= \Tilde{v}_{k} + \rho\big(o_{\mathrm{r},k+1}-v_{k+1}\big)\\[0.5em]
    \Tilde{w}_{k+1} &= \Tilde{w}_{k} + \rho\big(Do_{\mathrm{r},k+1}-w_{k+1}\big)
\end{align*}

Here, $o_\mathrm{r} \in \mathbb{R}^{n_x n_y n_z}$ is the vectorized reconstructed object, $b_\mathrm{m} \in \mathbb{R}^{n_x n_y n_z}_+$ is the vectorized measured data, $P\in \mathbb{R}^{n_x n_y n_z\times n_x n_y n_z}$ is the convolution matrix corresponding to the convolution with the PSF, $D\in \mathbb{R}^{n_x n_y n_z\times n_x n_y n_z}$ is the isotropic total-variation operator, $\Tilde{v}, \Tilde{w}\in \mathbb{R}^{n_x n_y n_z}$ are the dual variables of $v$ and $w$, and $\rho, \alpha \in \mathbb{R}$ are constants defining the convergence properties of the algorithm. 
To obtain our reconstructed image, we define the initial conditions $o_{r,0},v_0,w_0,\Tilde{v}_0,\Tilde{w}_0$ as the zero vectors and compute $k$ iterations until we converge to a solution.

\section{Appendix E: Sample topography} \label{APP:topography}
\begin{figure}[htb]
\includegraphics[width=\columnwidth]{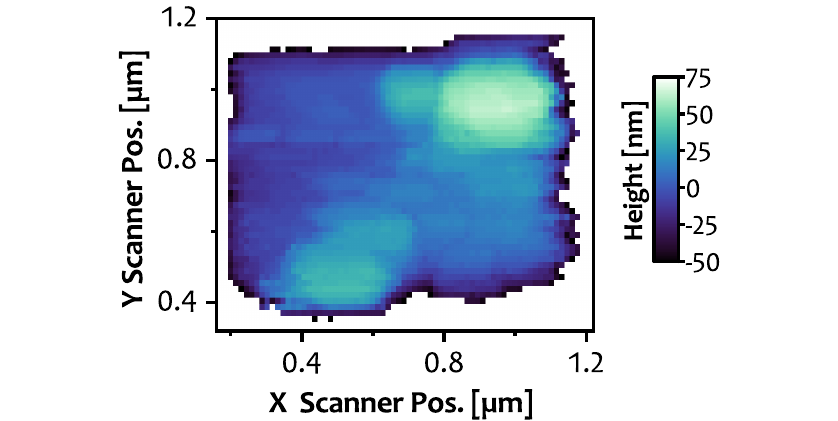}
\caption{Topography scan of the \SI{870}{\nano\meter} x \SI{630}{\nano\meter} rectangular cantilever apex with the attached virus samples. The scan was performed at room-temperature after performing the 3D MRFM scan.
}
\label{fig:figS4}
\end{figure}

A topography scan of the cantilever tip with the virus samples is shown in Fig.~\ref{fig:figS4}. The image was recorded by approaching a sharp cone fabricated on a commercial AFM tip calibration grid to the flat cantilever apex at different positions and recording the height at which a touch was registered. The criterion for a touch was defined here as either the oscillation frequency of the cantilever dropping below \SI{500}{\hertz} or rising above \SI{13500}{\hertz}. An alternative touch condition is the reduction of the amplitude below a threshold value. 

\bibliographystyle{apsrev4-1}
\bibliography{references}


\end{document}